\documentclass{elsart}

\usepackage{amssymb}
\usepackage{epsfig}
\usepackage{amsmath}
\usepackage{amsfonts}
\usepackage{epsfig}

\begin{document}

\begin{frontmatter}

% Title, authors and addresses

% use the thanksref command within \title, \author or \address for footnotes;
% use the corauthref command within \author for corresponding author footnotes;
% use the ead command for the email address,
% and the form \ead[url] for the home page:
% \title{Title\thanksref{label1}}
% \thanks[label1]{}
% \author{Name\corauthref{cor1}\thanksref{label2}}
% \ead{email address}
% \ead[url]{home page}
% \thanks[label2]{}
% \corauth[cor1]{}
% \address{Address\thanksref{label3}}
% \thanks[label3]{}

\title{Relativistic approach to electromagnetic imaging}

% use optional labels to link authors explicitly to addresses:
% \author[label1,label2]{}
% \address[label1]{}
% \address[label2]{}

\author{Neil V. Budko}
\ead{n.budko@ewi.tudelft.nl}
\address{Laboratory of Electromagnetic Research, Faculty of Electrical Engineering,
Mathematics and Computer Science, Delft University of Technology,
Mekelweg 4, 2628 CD Delft, The Netherlands}

\begin{abstract}
A novel imaging principle based on the interaction of electromagnetic waves
with a beam of relativistic electrons is proposed. Wave-particle interaction is
assumed to take place in a small spatial domain, so that each electron is only briefly accelerated
by the incident field. In the one-dimensional case the spatial distribution of the
source density can be directly observed in the temporal spectrum of the scattered field.
Whereas, in the two-dimensional case the relation between the source and 
the spectrum is shown to be approximately the Radon transform.
\end{abstract}

\end{frontmatter}

% main text
\section{Introduction}
\label{Intro}
%%%%%%%%%%%%%%%%%%%%%%%%%%%%%%%%%%%%%%%%%%%%%%%%%%%%%%%%%%%%%%%%%%%%%%%%%
Telescopes, antenna arrays, radars, and imaging interferometers share a common
resolution criterion: the spatial extent of their aperture.
Higher resolution means a larger lens, a larger reflector, or a longer interferometer base. 
At the present stage any significant improvement of the resolution 
is associated with extreme technical difficulties and costs, and for some applications 
we are approaching a limit in this respect. Hence, it makes sense 
to consider alternative imaging principles, especially ones where the attainable
resolution would not be entirely controlled by the spatial extent of some `aperture'.

In a variety of cases both the observer and the observed object are in relative motion
with respect to each other. A well-known phenomenon related to this motion
is the Doppler effect. It is widely exploited in radar \cite{Bertrand1997}, and 
acoustical imaging of blood vessels \cite{Sparr1995}, \cite{Schuster2000}, 
and is sometimes used to improve the resolution in 
astronomy \cite{Desbat1999}. Not to mention the famous red shift of
stars and galaxies around us, which is the main source of cosmological 
information. However, in all of these imaging techniques the observer is actually considered 
to be at rest. Often, this is just a question of reference frame, and we simply 
find it convenient to relate ourselves with the one at rest. 
If for some reason the motion of the observer cannot be neglected
in this way, then our natural intention is to compensate for it \cite{Huang2003},
\cite{Fridman2003}.

In this paper we investigate the possibility of exploiting the observer's motion
in electromagnetic imaging, rather than neglecting or correcting for it. 
Apparently, for this purpose we need an observer that moves with a relativistic velocity.
This, however, presents not a problem, since an electron, which is easily accelerated to such
velocities, is also the most natural ``observer'' for the electromagnetic field.
An obvious, but not at all unique way to organize an imaging system of this kind is to let a beam of 
relativistic electrons interact with the fields radiated by some remote 
spatially inhomogeneous source distribution, say a group of stars. 
Contrary to standard (stationary)
imaging instruments, here it is the observer's velocity, not the spatial extent of the aperture, that 
determines the attainable resolution.

One of the problems in \cite{LandauLifshitz} (p.~249),
where a plane monochromatic wave interacts with a single relativistic 
electron, contains a hint, which has inspired
the present paper. In \cite{LandauLifshitz} it is claimed that upon scattering another 
plane wave can be observed in the far-field, which has a different frequency with 
respect to the incident one, depending on the angles of incidence and observation. 
Here we simulate an imaging setup and, therefore, consider incident fields due to a 
remote but spatially extended source.
These fields interact with a short segment of a relativistic current.
The short-segment assumption allows to circumvent the difficult
problem of charge dynamics. Straightforward calculations presented below show 
that the relation between the spatial density distribution of the source and the 
(temporal) spectrum of the secondary field radiated by the electrons 
is approximately a Radon transform.
This transform is well known in imaging theory and is amenable to invertion \cite{NattererBook}. 
Mathematicaly, the particular form of the Radon transform
obtained here is very similar to the Doppler and Doppler-Radon transforms, which are extensively 
studied in the (non-relativistic) acoustical imaging of fluid flows \cite{Sparr1995}, \cite{Schuster2000}.

The author does not claim to have covered all physical and mathematical aspects of 
this interesting problem. In fact, to arrive at the Radon transform in its simplest form
we make several approximations, the significance of which must be further
investigated both theoretically and experimentally.

% The Appendices part is started with the command \appendix;
% appendix sections are then done as normal sections
% \appendix

% \section{}
% \label{}
\section{Theory}
%%%%%%%%%%%%%%%%%%%%%%%%%%%%%%%%%%%%%%%%%%%%%%%%%%%%%%%%%%%%%%%%%%%%%%%%%
Let there be a remote spatially inhomogeneous and time varying source of the 
electromagnetic field. We consider the scattering of this field by a 
segment of relativistic electric current, e.g. a beam of electrons moving with 
a relativistic velocity in a particle accelerator or even a simple cathode 
ray tube. Neglecting the mutual interaction between the electrons, quantum effects,
as well as the radiation reaction force, 
we would like to calculate the secondary electromagnetic field,
scattered by this relativistic current into the far-field zone. 

Let ${\mathbf x}_{\rm e}(t)$ denote the location of a single moving electron
and let $D_{\rm s}$ be a bounded region of space occupied 
by an external source of the electromagnetic field (see Fig.~1).
Presume that $D_{\rm s}$ is far enough both from the origin of the coordinate
system and location ${\mathbf x}_{\rm e}(t)$, so that the distance factor
can be approximated by
 \begin{align}
 \label{eq:Dist1}
 \vert{\mathbf x}_{\rm s} - {\mathbf x}_{\rm e}(t_{\rm e})\vert \approx 
 \vert{\mathbf x}_{\rm s}\vert
 -{\mathbf n}_{\rm s}\cdot{\mathbf x}_{\rm e}(t_{\rm e}), 
 \;\;\;\text{for}\;\;\; {\mathbf x}_{\rm s} \in D_{\rm s},
 \end{align}
where ${\mathbf n}_{\rm s}={\mathbf x}_{\rm s}/\vert{\mathbf x}_{\rm s}\vert$.
In Gaussian units the incident electromagnetic field at the location of an electron 
is given by
 \begin{align}
 \label{eq:Ein}
 {\mathbf E}^{\rm in}({\mathbf x}_{\rm e}, t_{\rm e})
 &=
 \frac{1}{c^{2}}
 \int\limits_{{\mathbf x}_{\rm s}\in D_{\rm s}}
 \frac{
 {\mathbf n}_{\rm s}\times\left({\mathbf n}_{\rm s}\times
 \partial_{t_{\rm s}}{\mathbf J}({\mathbf x}_{\rm s},t_{\rm s})\right)}
 {\vert{\mathbf x}_{\rm s}\vert} \;{\rm d} V_{\rm s},
 \\
 \label{eq:Hin}
 {\mathbf H}^{\rm in}({\mathbf x}_{\rm e}, t_{\rm e})
 &=
 \frac{1}{c^{2}}\int\limits_{{\mathbf x}_{\rm s}\in D_{\rm s}}
 \frac{{\mathbf n}_{\rm s}\times\partial_{t_{\rm s}}{\mathbf J}({\mathbf x}_{\rm s},t_{\rm s})}
 {\vert{\mathbf x}_{\rm s}\vert} \;{\rm d} V_{\rm s},
 \end{align}
where ${\mathbf J}({\mathbf x},t)$ is the electric current density in the remote source.
Time $t_{\rm s}$ in the right-hand-sides of (\ref{eq:Ein})--(\ref{eq:Hin}) 
is retarded with respect to time $t_{\rm e}$.

\begin{figure}[t]
\centerline{\epsfig{file=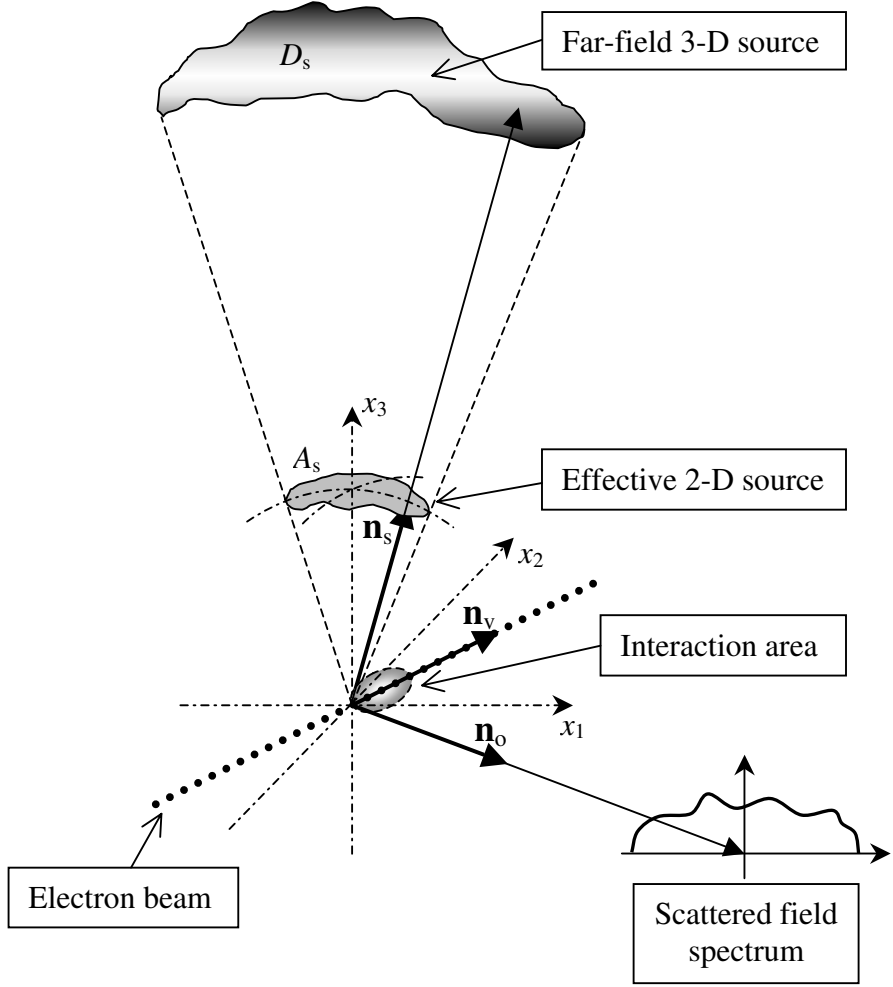,width=0.9\columnwidth,height=\columnwidth}}
\caption{Imaging setup with a relativistic detector (electron beam).}
\end{figure}

The acceleration of a single electron due to these fields is
given by the well-known relativistic formula \cite{LandauLifshitz}
 \begin{align}
 \label{eq:Acceleration}
 \begin{split}
 \partial_{t_{\rm e}}{\mathbf v}_{\rm e} =-
 \frac{e}{m_{\rm e}}\sqrt{1-\beta^{2}}
 &\left[{\mathbf E}^{\rm in}({\mathbf x}_{\rm e},t_{\rm e})- 
 \beta^{2}{\mathbf n}_{\rm v}\left({\mathbf n}_{\rm v}\cdot{\mathbf E}^{\rm in}({\mathbf x}_{\rm e},t_{\rm e})\right) 
 \right.
 \\
 &\left.
 +\beta\, {\mathbf n}_{\rm v}\times{\mathbf H}^{\rm in}({\mathbf x}_{\rm e},t_{\rm e})
 \right],
 \end{split}
 \end{align}
where $\beta=\vert{\mathbf v}_{\rm e}\vert/c$, ${\mathbf n}_{\rm v}
={\mathbf v}_{\rm e}/\vert{\mathbf v}_{\rm e}\vert$, and ${\mathbf v}_{\rm e}$
is the electron velocity. 
This formula clearly shows that we need to take into account the action of both the 
electric and the magnetic fields
of the source.
Subsequently, an electron radiates into 
the far-field the following electric field \cite{LandauLifshitz}:
 \begin{align}
 \label{eq:Esc}
 {\mathbf E}({\mathbf x}_{\rm o},t_{\rm o})
 =-
 \frac{e}{c^{2}\vert{\mathbf x}_{\rm o}\vert\;
 [1-\beta ({\mathbf n}_{\rm o}\cdot{\mathbf n}_{\rm v})]^{3}}
 {\mathbf n}_{\rm o}\times
 \left[
 \left({\mathbf n}_{\rm o} - \beta {\mathbf n}_{\rm v} \right) 
 \times \partial_{t_{\rm e}}{\mathbf v}_{\rm e}
 \right].
 \end{align}
Here, ${\mathbf x}_{\rm o}$ denotes the location of the observation 
point. Time $t_{\rm e}$ in the right-hand-side of this 
expression is retarded with respect to time $t_{\rm o}$.
Upon substituting (\ref{eq:Ein})--(\ref{eq:Hin}) in (\ref{eq:Acceleration}),
and the result in (\ref{eq:Esc}), we arrive at
 \begin{align}
 \label{eq:EviaJ}
 \begin{split}
 {\mathbf E}({\mathbf x}_{\rm o},t_{\rm o})
 &=
 \frac{e^{2}\,\sqrt{1-\beta^{2}}}
 {c^{4} m_{\rm e}\vert{\mathbf x}_{\rm o}\vert\,\left[1-\beta\,({\mathbf n}_{\rm o}\cdot{\mathbf n}_{\rm v})\right]^{3}}
 \\
 {\mathbf n}_{\rm o}\times
 &\left\{
 \left[{\mathbf n}_{\rm o} - \beta {\mathbf n}_{\rm v} \right] 
 \times 
 \left[
 \int\limits_{\;\;{\mathbf x}_{\rm s}\in D_{\rm s}}
  \frac{
  {\mathbf n}_{\rm s}\times\left({\mathbf n}_{\rm s}\times
  \partial_{t_{\rm s}}{\mathbf J}({\mathbf x}_{\rm s},t_{\rm s})\right)}
 {\vert{\mathbf x}_{\rm s}\vert} \;{\rm d} V_{\rm s}
 -
 \right.
 \right.
 \\
 &\;\;
 \beta^{2}{\mathbf n}_{\rm v}\left({\mathbf n}_{\rm v}\cdot
 \int\limits_{{\mathbf x}_{\rm s}\in D_{\rm s}}
 \frac{
 {\mathbf n}_{\rm s}\times\left({\mathbf n}_{\rm s}\times
 \partial_{t_{\rm s}}{\mathbf J}({\mathbf x}_{\rm s},t_{\rm s})\right)}
 {\vert{\mathbf x}_{\rm s}\vert} \;{\rm d} V_{\rm s}
 \right)
 \\
 &\;\;
 \left.
 \left.
 +
 \beta\,{\mathbf n}_{\rm v}\times
 \int\limits_{{\mathbf x}_{\rm s}\in D_{\rm s}}
  \frac{{\mathbf n}_{\rm s}\times\partial_{t_{\rm s}}{\mathbf J}({\mathbf x}_{\rm s},t_{\rm s})}
 {\vert{\mathbf x}_{\rm s}\vert} \;{\rm d} V_{\rm s}
 \right]
 \right\}
 \end{split}
 \end{align}
This expression describes the scattering by a single electron. 
The electric field scattered by a beam of 
non-interacting electrons is simply the sum of the fields scattered by each 
electron individually. All these fields can be computed using (\ref{eq:EviaJ}),
provided that the electron velocities are given. 
In principle, these velocities are solutions of the relativistic equation 
of motion (\ref{eq:Acceleration}), which is a nonlinear equation and therefore 
very difficult, if not impossible, to solve.
For our purposes, however, it is not necessary to know
the velocities exactly. We rather need a reliable estimate on their
variation with respect to the initial relativistic velocity. 

To obtain such an estimate we use the Taylor expansion
 \begin{align}
 \label{eq:VelocityExp}
 {\mathbf v}_{\rm e}(t_{0}+\Delta t) = {\mathbf v}_{\rm e}(t_{0}) + 
 \partial_{t}{\mathbf v}_{\rm e}\vert_{t=t_{0}}\Delta t + \text{higher order terms},
 \end{align}
and observe that ${\mathbf v}_{\rm e}(t)$ can be approximated by the 
initial velocity up to the order $\Delta t$ -- the time 
of the field-particle interaction. Note, if $\Delta t$ 
is sufficiently small, then the initial velocity approximation is suitable even for nonvanishing accelerations,
and is, therefore, consistent with the idea of the secondary (i.e. acceleration related) radiation.
Since $\Delta t$ is approximately the time of flight of
electrons across the interaction area, one can decrease $\Delta t$ by decreasing this 
area. This can be achieved by either 
actually creating a relatively short relativistic current 
(radio frequencies) or by focusing the incident fields 
(optics), so that the latter are different from zero only within a certain 
finite and relatively small spatial domain. 
Under these assumtions we may take in (\ref{eq:EviaJ})
 \begin{align}
 \label{eq:VelocityApp1}
 \begin{split}
 {\mathbf v}_{\rm e}(t_{\rm e})&\approx{\mathbf v}_{\rm e}(t_{0})={\rm constant},
 \\
 \beta(t_{\rm e})&\approx\beta(t_{0})= {\rm constant},
 \\
 {\mathbf n}_{\rm v}(t_{\rm e})&\approx{\mathbf n}_{\rm v}(t_{0})={\rm constant}.
 \end{split}
 \end{align}
As far as the electron location ${\mathbf x}_{\rm e}(t_{\rm e})$ is concerned, 
we are only interested in ${\mathbf x}_{\rm e}(t_{\rm e})$ within the (small) 
area of interaction. For each electron this location depends on time linearly 
up to ${\mathcal O}(\Delta t)$. Under these assumptions the distance expression (\ref{eq:Dist1}) becomes
 \begin{align}
 \label{eq:DistSlin}
 \vert{\mathbf x}_{\rm s}\vert
 -{\mathbf n}_{\rm s}\cdot{\mathbf x}_{\rm e}(t_{\rm e}) 
 \approx
 \vert{\mathbf x}_{\rm s}\vert
 -t_{\rm e}\, ({\mathbf n}_{\rm s}\cdot{\mathbf v}_{\rm e}) + 
 {\mathbf n}_{\rm s}\cdot{\mathbf x}_{\rm e}(t_{0}),
 \end{align}
where ${\mathbf x}_{\rm e}(t_{0})$  is the initial location of the electron inside 
the area.

We set the origin of the coordinate system to be within the area of interaction.  
Recall that the times at the source and the electron locations are retarded 
with respect to each other and with respect to the time at the observation point. 
In the initial velocity approximation these times are
 \begin{align}
 \label{eq:TimeS}
 t_{\rm s} & = t_{\rm e} - \frac{\vert{\mathbf x}_{\rm e}(t_{\rm e})-{\mathbf x}_{\rm s}\vert}{c}
 \approx
 t_{\rm e}\left[1 + \beta ({\mathbf n}_{\rm s}\cdot{\mathbf n}_{\rm v})\right] - \frac{\vert{\mathbf x}_{\rm s}\vert}{c},
 \\
 \label{eq:TimeE}
 t_{\rm e} & = t_{\rm o} - \frac{\vert{\mathbf x}_{\rm o}-{\mathbf x}_{\rm e}(t_{\rm e})\vert}{c}
 \approx
 t_{\rm o} + t_{\rm e}\,\beta ({\mathbf n}_{\rm o}\cdot{\mathbf n}_{\rm v})
 - \frac{\vert{\mathbf x}_{\rm o}\vert}{c}.
 \end{align}
up to the order $\Delta t$. The arbitrary initial locations of electrons vanish, 
and, therefore, approximately the same time is now associated with all electrons inside 
the interaction area.
From (\ref{eq:TimeS})--(\ref{eq:TimeE}) we obtain the following 
relation between the times at the source location and the observation point:
 \begin{align}
 \label{eq:TimeO}
 t_{\rm s}=\left(t_{\rm o}-\frac{\vert{\mathbf x}_{\rm o}\vert}{c}\right)
 \frac{1+\beta ({\mathbf n}_{\rm s}\cdot{\mathbf n}_{\rm v})}
 {1-\beta ({\mathbf n}_{\rm o}\cdot{\mathbf n}_{\rm v})} - \frac{\vert{\mathbf x}_{\rm s}\vert}{c}.
 \end{align}

A particularly simple expression in the case of almost constant 
velocities is obtained for the following choice of the observation direction:
 \begin{align}
 \label{eq:NoNv}
 \begin{split}
 {\mathbf n}_{\rm o}&={\mathbf n}_{\rm v},
 \\
 {\mathbf n}_{\rm o}\cdot{\mathbf n}_{\rm v}&=1.
 \end{split}
 \end{align}
This choice reduces equation (\ref{eq:EviaJ}) to
 \begin{align}
 \label{eq:EviaJDir}
 \begin{split}
 {\mathbf E}({\mathbf x}_{\rm o},t_{\rm o})
 &=
 \frac{N e^{2}\,\sqrt{1-\beta^{2}}}
 {c^{4}m_{\rm e}\vert{\mathbf x}_{\rm o}\vert\,\left(1-\beta\right)^{2}}
 \\
 &
 \left\{
 {\mathbf n}_{\rm v}\times\left[{\mathbf n}_{\rm v}\times
 \hspace*{-0.3cm}
 \int\limits_{\;\;{\mathbf x}_{\rm s}\in D_{\rm s}}
 \frac{{\mathbf n}_{\rm s}\times\left({\mathbf n}_{\rm s}\times
 \partial_{t_{\rm s}}{\mathbf J}({\mathbf x}_{\rm s},t_{\rm s})\right)}
 {\vert{\mathbf x}_{\rm s}\vert} \;{\rm d} V_{\rm s}
 \right]
 \right.
 \\
 &\;
 \left.
 +
 \beta\,{\mathbf n}_{\rm v}\times
 \hspace*{-0.3cm}
 \int\limits_{{\mathbf x}_{\rm s}\in D_{\rm s}}
 \frac{{\mathbf n}_{\rm s}\times\partial_{t_{\rm s}}{\mathbf J}({\mathbf x}_{\rm s},t_{\rm s})}
 {\vert{\mathbf x}_{\rm s}\vert} \;{\rm d} V_{\rm s}
 \right\},
 \end{split}
 \end{align}
where $N$ is the total number of electrons within the interaction area, which, presumably,
is the same at any time instant.

Consider a single harmonic component of the time derivative of the source current density
 \begin{align}
 \label{eq:HarmJ}
 i\omega_{\rm s}\hat{\mathbf J}({\mathbf x}_{\rm s},\omega_{\rm s})\exp(i\omega_{\rm s}t_{\rm s})
 =
 i\omega_{\rm s}\hat{\mathbf J}({\mathbf x}_{\rm s},\omega_{\rm s})
 \exp\left(-i\frac{\omega_{\rm s}}{c}\vert{\mathbf x}_{\rm s}\vert\right)
 \exp\left(i\omega t\right),
 \end{align}
where
 \begin{align}
 \label{eq:Omega}
 \omega&=
 \omega_{\rm s}\frac{1+\beta({\mathbf n}_{\rm s}\cdot{\mathbf n}_{\rm v})}
 {1-\beta({\mathbf n}_{\rm o}\cdot{\mathbf n}_{\rm v})}
 = \omega_{\rm s}\frac{1+\beta({\mathbf n}_{\rm s}\cdot{\mathbf n}_{\rm v})}
 {1-\beta},
 \\
 \label{eq:T}
 t&=t_{\rm o}-\frac{\vert{\mathbf x}_{\rm o}\vert}{c}.
 \end{align}
Remarkably, frequency relation (\ref{eq:Omega}) coincides with the one
obtained in \cite{LandauLifshitz} for the plane-wave case, which did not involve any
approximations. We see now that even to a purely harmonic but spatially extended source
there corresponds a whole set of frequencies in the scattered wave, which depend
on the particular values admitted by ${\mathbf n}_{\rm s}$ and ${\mathbf n}_{\rm v}$
in (\ref{eq:Omega}). To obtain a more elaborate relation between the spatial 
distribution of the source current density and the spectrum of the scattered wave
we take the Fourier transform of the observed electric field strength
with respect to time. But first we simplify (\ref{eq:EviaJDir}) by resorting to 
the effective two-dimensional aperture $A_{\rm s}$ (see Fig.~1), so that for a purely
harmonic source (\ref{eq:HarmJ}) we have
 \begin{align}
 \label{eq:EviaehDir}
 \begin{split}
 {\mathbf E}({\mathbf n}_{\rm v},t_{\rm o})
 =
 C
 \int\limits_{{\mathbf n}_{\rm s} \in A_{\rm s}}
 \left\{
 {\mathbf n}_{\rm v}\times\left[{\mathbf n}_{\rm v}\times
 {\mathbf e}({\mathbf n}_{\rm s},\omega_{\rm s})\right]
 \right.
 \\
 \left.
 +
 \beta\,\left[{\mathbf n}_{\rm v}\times
 {\mathbf h}({\mathbf n}_{\rm s},\omega_{\rm s})\right]
 \right\}
 \exp\left(i\omega t\right)
 \;{\rm d} A\,,
 \end{split}
 \end{align}
where ${\mathbf e}$ and ${\mathbf h}$ are the effective transverse fields with 
inhomogeneous angular distribution. All unimportant coefficients here and below are
lumped with the constant of proportionality $C$. Notice that factor $\exp\left(i\omega t\right)$
cannot be taken outside the integral, since $\omega$ depends on the integration
variable. The Fourier transform of the left-hand-side must be taken with
respect to the observation time. Whereas, using the following simple
manipulation:
 \begin{align}
 \label{eq:FourierT}
 \begin{split}
 \hat{\mathbf E}({\mathbf n}_{\rm v},\omega_{\rm o})=
 \int\limits_{-\infty}^{\infty}{\mathbf E}({\mathbf n}_{\rm v},t_{\rm o})\exp\left(-i\omega_{\rm o} t_{\rm o}\right) \;{\rm d} t_{\rm o} 
 =
 \\
 \exp\left(-i\frac{\omega_{\rm o}}{c}\vert{\mathbf x}_{\rm o}\vert\right)
 \int\limits_{-\infty}^{\infty}{\mathbf E}\left({\mathbf n}_{\rm v},t+\frac{\vert{\mathbf x}_{\rm o}\vert}{c}\right)
 \exp\left(-i\omega_{\rm o} t\right) \;{\rm d} t,
 \end{split}
 \end{align}
in the right-hand-side we may take the Fourier transform with respect to retarded time $t$.
The observed frequency is denoted $\omega_{\rm o}$. Now we substitute
(\ref{eq:EviaehDir}) in (\ref{eq:FourierT}) and arrive at
 \begin{align}
 \label{eq:EFT}
 \begin{split}
 \hat{\mathbf E}({\mathbf n}_{\rm v},\omega_{\rm o})
 =
 C
 \int\limits_{{\mathbf n}_{\rm s} \in A_{\rm s}}
 \left\{
 {\mathbf n}_{\rm v}\times\left[{\mathbf n}_{\rm v}\times
 {\mathbf e}({\mathbf n}_{\rm s},\omega_{\rm s})\right]
 \right.
 \\
 \left.
 +
 \beta\,\left[{\mathbf n}_{\rm v}\times
 {\mathbf h}({\mathbf n}_{\rm s},\omega_{\rm s})\right]
 \right\}
 \delta\left(\omega-\omega_{\rm o}\right)
 \;{\rm d} A\,,
 \end{split}
 \end{align}
where $\delta\left(\omega-\omega_{\rm o}\right)$ is the Dirac delta-function,
which for $\beta\ne 0$, and $\omega_{\rm s}\ne 0$, can be alternatively expressed as
 \begin{align}
 \label{eq:DeltaOmega}
 \begin{split}
 \delta\left(\omega-\omega_{\rm o}\right)=
 &\;\delta\left[ \omega_{\rm s}\frac{1+\beta({\mathbf n}_{\rm s}\cdot{\mathbf n}_{\rm v})}
 {1-\beta}-\omega_{\rm o}\right]
 \\
 =&\;\frac{1 - \beta}
 {\beta\vert\omega_{\rm s}\vert}
 \;\delta\left[
 \frac{\omega_{\rm o}(1 - \beta)-\omega_{\rm s}}
 {\beta\omega_{\rm s}}-({\mathbf n}_{\rm s}\cdot{\mathbf n}_{\rm v})\right].
 \end{split}
 \end{align}
Upon substitution of this expression in (\ref{eq:EFT}), similarity with the Radon transform
or, more precisely, the Doppler transform \cite{Sparr1995}, \cite{Schuster2000} becomes obvious. 
The vectorial structure of equation (\ref{eq:EFT}) 
is still quite complex, however. The intensity of harmonics
looks somewhat simpler:
 \begin{align}
 \label{eq:Intensity}
 \begin{split}
 \left(\overline{\hat{\mathbf E}}\cdot{\hat{\mathbf E}}\right)=&\;
 \left(\overline{\mathbf I}_{\rm e}\cdot{\mathbf I}_{\rm e}\right)
 +\beta^{2}\left(\overline{\mathbf I}_{\rm h}\cdot{\mathbf I}_{\rm h}\right)
 \\
 &-\left\vert\left({\mathbf n}_{\rm v}\cdot{\mathbf I}_{\rm e}\right)\right\vert^{2}
 -\beta^{2}\left\vert\left({\mathbf n}_{\rm v}\cdot{\mathbf I}_{\rm h}\right)\right\vert^{2}
 +2\beta\left({\mathbf n}_{\rm v}\cdot{\rm Re}\left\{\overline{\mathbf I}_{\rm e}\times{\mathbf I}_{\rm h}\right\}\right),
 \end{split}
 \end{align}
where overbar denotes complex conjugation, and
 \begin{align}
 \label{eq:IntegralE}
 {\mathbf I}_{\rm e}&=
  C
  \int\limits_{{\mathbf n}_{\rm s} \in A_{\rm s}}
  {\mathbf e}({\mathbf n}_{\rm s},\omega_{\rm s})
  \delta\left[\omega'-({\mathbf n}_{\rm s}\cdot{\mathbf n}_{\rm v})\right]
  \;{\rm d} A\,,
 \\
 \label{eq:IntegralH}
 {\mathbf I}_{\rm h}&=
 C
 \int\limits_{{\mathbf n}_{\rm s} \in A_{\rm s}}
 {\mathbf h}({\mathbf n}_{\rm s},\omega_{\rm s})
 \delta\left[\omega'-({\mathbf n}_{\rm s}\cdot{\mathbf n}_{\rm v})\right]
 \;{\rm d} A\,. 
 \end{align}
Splitting these up into components orthogonal and parallel to ${\mathbf n}_{\rm v}$,
i.e.
 \begin{align}
 \label{eq:Split}
 {\mathbf I}_{\rm e,h}={\mathbf I}_{\rm e,h}^{\parallel}+{\mathbf I}_{\rm e,h}^{\perp},
 \end{align}
we finally obtain
 \begin{align}
 \label{eq:IntensitySplit}
 \begin{split}
 \left(\overline{\hat{\mathbf E}}\cdot{\hat{\mathbf E}}\right)=&\;
 \left(\overline{{\mathbf I}_{\rm e}^{\perp}}\cdot{\mathbf I}_{\rm e}^{\perp}\right)
 +\beta^{2}\left(\overline{{\mathbf I}_{\rm h}^{\perp}}\cdot{\mathbf I}_{\rm h}^{\perp}\right).
 \end{split}
 \end{align}
Subsequent transformations involve
assumptions about the spatial coherence of the source, which are
beyond the scope of this paper, but can be found for example in \cite{Born&Wolf}. 

Anyway, it is clear that, if we consider two incoherent point 
sources, then 
the observed field has only two Fourier components corresponding
to the frequencies, which can be easily determined from (\ref{eq:Omega}).
In other words, we shall directly observe an image of our sources in the
temporal spectrum of the field. Separation between the spectral lines 
depends on the angular separation of the sources and the electron 
velocity. The higher the velocity, the better the attainable 
spatial (angular) resolution. 

\section{Discussion}
%%%%%%%%%%%%%%%%%%%%%%%%%%%%%%%%%%%%%%%%%%%%%%%%%%%%%%%%%%%%%%%%%%%%%%%%%

Unfortunately, there is no such thing in Nature as a 
purely monochromatic source of type (\ref{eq:HarmJ}). Therefore, even in the preceding simple one-dimensional 
example one should expect broadening of each 
of the two observed lines proportional to the broadening of the source's spectral 
line. In terms of the image quality this leads to the loss of spatial resolution --
blurring. In addition, spectral resolution of the
dispersing instrument (e.g. a prism or a grating) must be taken into account.
Imagine two identical dispersing instruments. We place one of them
on the path between the source and the interaction area, so that only one
spectral line of width $\Delta\omega_{\rm s}$ is allowed through.
Another dispersing instrument is used to investigate the
spectral content of the scattered field.
Then, for the two sources to be resolved, the electron velocity must satisfy 
 \begin{align}
 \label{eq:Resolution}
 \frac{\vert{\mathbf v}_{\rm e}\vert}{c}\ge \frac{1}{1+R\,\Delta\varphi},
 \end{align}
where $R=\omega_{\rm o}/\vert\Delta\omega_{\rm o}\vert$ is, by definition, the resolving power of a spectroscope, 
and $\Delta\varphi=\vert\sin\varphi_{1}-\sin\varphi_{2}\vert$ is the angular separation of the sources
(the sources and the electron beam are coplanar).

Another anticipated cause of blurring is the variation 
of the particle velocity inside the interaction area, if the latter could not be made 
sufficiently small. The Fourier transform of (\ref{eq:EviaehDir}) 
with a time-varying velocity will then introduce an extra convolution in $\omega_{\rm o}$-domain
with the temporal Fourier transform of the electron velocity. 

Correspondence between the spectrum and the source for 
two-dimensional distributions is not as direct as in the one-dimensional case. 
To invert the Radon transform spectral data for different directions
of the velocity vector must be collected, i.e. we have to introduce the mutual 
rotation between the source and the particle accelerator around the axis perpendicular 
to the plane of the source. Subsequently, numerical inversion of the Radon
transform must be carried out.

In summary, the proposed relativistic alternative to aperture-based imaging
consists of employing a beam of relativistic electrons as a detector
in an optical imaging system. 
A classical treatment of this problem has been presented. A more refined analysis would have to
take into account: the mutual interaction between electrons, the radiation reaction force,
and the quantum effects. The question of coherence 
must be investigated in more detail as well. 

\section*{Acknowledgement}
The author is grateful to Prof.~A.~T.~de~Hoop (Delft University of Technology) 
for his help and support. In particular, for his advice to tackle the problem using the 
relativistic equation of motion. 
The author also appreciates numerous discussions with Dr.~R.~Remis (Delft University of Technology).

\end{document}